\documentclass[superscriptaddress,twocolumn,aps,prl,floatfix,showpacs]{revtex4}
\usepackage{epsfig}
\usepackage{dcolumn}
\usepackage{bm}
\begin{document}

\title{Universal Behavior of the Resistance Noise Across
the Metal-Insulator Transition in Silicon Inversion Layers}
\author{J.\ Jaroszy\'nski}
\email{jaroszy@magnet.fsu.edu} 
\altaffiliation[also at ]{Institute of
Physics, PAS, Warsaw, Poland}
\affiliation{National High Magnetic Field Laboratory, Florida State
University, Tallahassee, Florida 32310}
\author{Dragana  Popovi\'c}
\affiliation{National High Magnetic Field Laboratory, Florida State
University, Tallahassee, Florida 32310}
\author{T.\ M.\ Klapwijk}
\affiliation{Department of Applied Physics, Delft University of
Technology, 2628 CJ Delft, The Netherlands} 
\date{\today}

\begin{abstract}
Studies of low-frequency resistance noise show that the glassy freezing
of the two-dimensional (2D) electron system in the vicinity of the 
metal-insulator transition occurs in all Si inversion layers.  
The size of the metallic glass phase, which separates the 2D metal 
and the (glassy) insulator, depends strongly on disorder, becoming extremely 
small in high-mobility samples.  The behavior of the second spectrum, an 
important fourth-order noise statistic, indicates the presence of long-range
correlations between fluctuators in the glassy phase, consistent with the 
hierarchical picture of glassy dynamics.
\end{abstract}


\maketitle

Despite many theoretical and experimental efforts, the metal-insulator 
transition \cite{SAK2000} (MIT) in two-dimensional (2D) systems remains 
controversial.  Since the apparent MIT occurs in the regime where both 
electron-electron interactions and disorder are strong, it has been suggested
that the 2D system undergoes glassy ordering in the vicinity of the MIT.  The 
proposals include freezing into a Coulomb \cite{thakur,PastorD99}, Wigner 
\cite{sudip}, or spin glass \cite{Sachdev}.  Indeed,
a recent study of low-frequency resistance noise in an extremely low-mobility
(high disorder) 2D electron system in Si demonstrated \cite{Bogd2002} glassy 
freezing, which occurred in the metallic phase as a precursor to the MIT.  

Here we report a detailed study of low-frequency resistance noise in a 2D 
electron system (2DES) in Si in the opposite limit of very low disorder, where
the 
metallic drop of resistivity $\rho$ with decreasing temperature $T$ is most 
pronounced.  Such samples have been studied extensively 
\cite{Heemskerk98,SAK2000} in the context of a 2D MIT using magnetotransport 
measurements.  We find that, similar to the case of low-mobility samples, the 
behavior of several spectral characteristics of noise in these high-mobility 
devices indicates a sudden and dramatic slowing down of the electron dynamics 
at a well-defined electron density $n_s=n_g$, corresponding to the transition 
to a glassy phase.  Since the two sets of devices, Si metal-oxide-semiconductor
field-effect transistors (MOSFETs), differ considerably by their peak 
mobility, which is a rough measure of the disorder, and span essentially the
entire range of Si technology, we conclude that the 
observed glass transition is a universal phenomenon in Si inversion layers.  
The experiments, however, have also revealed an important difference between 
low- and high-mobility samples.  In low-mobility devices, $n_g\approx 1.5\, 
n_c$ \cite{Bogd2002}, where $n_c$ is the critical density for the MIT 
determined from the vanishing of activation energy \cite{activated}, and the
temperature coefficient of $\rho$ changes sign at $n_{s}^{\ast}>n_g>n_c$.  In
high-mobility structures, on the other hand, the onset of glassy dynamics seems
almost to coincide with the MIT, {\textit i.~e.} $n_g\approx n_c\approx 
n_{s}^{\ast}$.  It is interesting 
that the observed strong dependence on disorder of the size of the metallic 
glass phase ($n_c<n_s<n_g$), which separates the 2D metal and the (glassy) 
insulator, is consistent with recent predictions of the model of interacting
electrons near a disorder-driven MIT~\cite{PastorD02}.  Furthermore,
by analyzing the second spectrum~\cite{Weissman88,Weissman93}, an 
important fourth-order noise statistic, we have established the presence of 
long-range correlations between fluctuators in the glassy phase, which provides
an unambiguous evidence for the onset of glassy dynamics at $n_g$.  The 
results are consistent with the picture in which noise, in the glassy phase, 
results from transitions between many metastable states with a hierarchical 
structure~\cite{Ogielski} with each transition being a reconfiguration of a 
large number of electrons.

\begin{figure}
\centerline{\epsfig{file=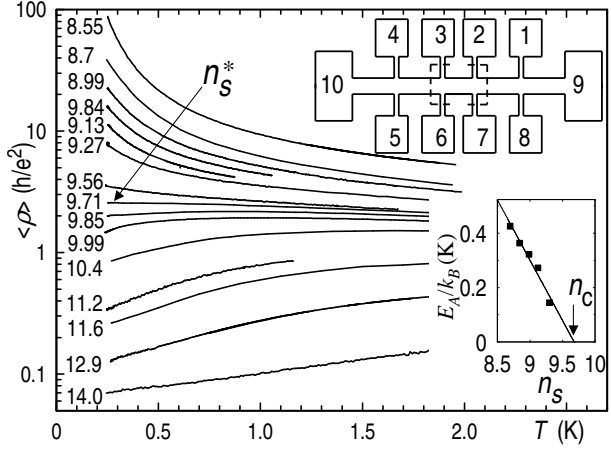,width=8cm,clip=}} \caption{
$\langle\rho\rangle$ {\textit vs.} $T$ for $n_s(10^{10}$cm$^{-2})$ shown on the
plot. Insets: sample schematic (dashed lines represent gaps in the 
gate), and activation energies {\textit vs.} $n_s$; $n_c\approx n_{s}^{\ast}$.
\label{average}}
\end{figure}

Measurements were carried out on n-channel Si MOSFETs with the peak mobility 
$\mu\approx 2.5$~m$^2$/Vs at 4.2~K, fabricated in a Hall bar geometry with Al 
gates, and oxide thickness $d_{ox}=147$~nm \cite{Heemskerk98}.  The resistance
$R$ was measured down to $T=0.24$~K using a standard four-probe ac technique
(typically $2.7$~Hz) in the Ohmic regime. A precision DC voltage standard 
(EDC MV116J) was used to apply the gate voltage, which controls $n_s$.  
Contact resistances and their influence on noise measurements were minimized 
by using a split-gate geometry, which allows one to maintain high $n_{s}$ 
($\approx10^{12}$~cm$^{-2}$) in the contact region while allowing an 
independent control of $n_s$ of the 2D system under investigation in the 
central part of the sample ($120\times 50~\mu$m$^2$) (Fig.~\ref{average} 
inset). 
%
%
Nevertheless, care was taken to ensure that the observed noise did not come 
from either the current contacts or the regions of gaps in the gate.
For example, since the noise measured across a resistor connected in series 
with the sample and having a similar resistance was at least three times lower
than the noise from the central part of the sample, the effect of the contact 
noise on the excitation current $I_{exc}$ could be easily ruled out.  
Similarly, the resistance and the noise measured between the voltage contact 
in the region of high $n_s$ ({\textit e.~g.} \#5 in Fig.~\ref{average} inset) 
and the one in the central part (\#6) were much smaller than those measured 
between contacts in the central part ({\textit e.~g.} \#6 and \#7).  In fact, 
they were in agreement with what is expected based on the geometry of the 
sample, which proves that the submicron gap regions did not contribute to 
either the measured resistance or noise.
In order to minimize the influence of fluctuations of both $I_{exc}$ and $T$, 
some of the noise measurements were carried out with a bridge configuration 
\cite{Scol87}.  The difference voltage was detected using two PAR124A lock-in 
amplifiers, and a cross-spectrum measurement was performed with an HP35665A 
spectrum analyzer in order to reduce the background noise even further 
\cite{Verbruggen89}.  The output filters of the lock-in amplifiers and/or 
spectrum analyzer served as an antialiasing device.  Most of the noise spectra
were obtained in the $f=(10^{-4}-10^{-1})$~Hz bandwidth, where the upper bound
was set by the low frequency of $I_{exc}$, limited by the low cut-off 
frequency of RC filters used to reduce external electromagnetic noise as well 
as by high $R$ of the sample.

Fig.~\ref{average} shows the time-averaged resistivity $\langle\rho\rangle$ as
a function of $T$ for different $n_s$; $d\langle\rho\rangle/dT=0$ at 
$n_{s}^{\ast}\approx 9.7\times 10^{10}$cm$^{-2}$.  For the lowest $n_s$ and
$T$, the data are described by the simply activated form
$\langle\rho\rangle\propto\exp (E_{A}/k_{B}T)$.  The vanishing of $E_A$ is 
often used as a criterion to determine $n_c$~\cite{activated}.  This method 
(Fig.~\ref{average} inset) yields $n_c\approx n_{s}^{\ast}$, in agreement with
other studies~\cite{activated}.

\begin{figure}
\centerline{\epsfig{file=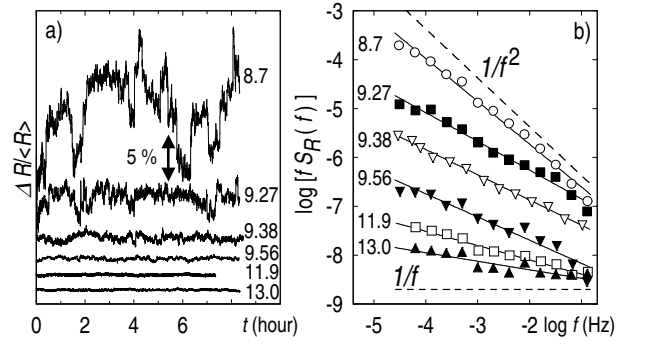,width=8cm,clip=}} 
\caption{(a) $\Delta R/\langle R\rangle$, and (b) the corresponding power 
spectra $S_{R}(f)$, for $n_s(10^{10}$cm$^{-2})$ shown on the plots; $T=0.24$~K.
In (a) traces 
are shifted for clarity.  In (b) $S_R(f)$ are averaged over octaves and 
multiplied by $f$, so that $1/f$ spectrum is horizontal on this scale.  Solid 
lines are linear least-squares fits.
\label{spectra1}}
\end{figure}

\begin{figure}
\centerline{\epsfig{file=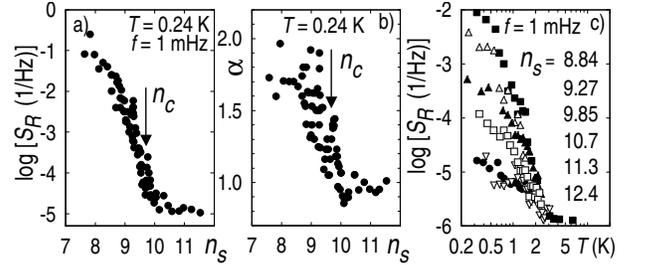,width=8cm,clip=}}
\caption{$S_R(f)$ {\textit vs.} $n_s (10^{10}$cm$^{-2})$ (a), and {\textit 
vs.} $T$ (c).  (b) $\alpha$ {\textit vs.} $n_s$.  $S_R(f)$ has been corrected 
for the white background noise.  The critical density $n_c$ is shown in (a) 
and (b).  
\label{expalpha}}
\end{figure}

Figures~\ref{spectra1}(a) and (b) show the time series of the relative changes
in resistance $\Delta R(t)/\langle R\rangle$,
%
%
and the corresponding power spectra $S_R(f)\propto 1/f^{\alpha}$, 
respectively.  In order to 
compare the noise magnitudes under different conditions, the spectra were 
averaged over two octaves [$(0.5-2)\times 10^{-3}$~Hz] around $f=10^{-3}$~Hz.
The resulting fraction of power $S_R(\mbox{$f=1$~mHz})$ is taken as the 
measure of noise, and its dependence on $n_s$ is shown in 
Fig.~\ref{expalpha}(a) for
%
%
$T=0.24$~K. The exponent $\alpha$ is displayed in 
Fig.~\ref{expalpha}(b), while Fig.~\ref{expalpha}(c) shows the dependence of 
$S_R(f=1$~mHz) on $T$ for several $n_s$.  Below $T\approx 3$~K, the noise 
increases with decreasing $T$, but at high $n_s$ where 
$d\langle\rho\rangle/dT>0$, $S_R$ depends rather weakly on both $T$ and $n_s$. 
In the vicinity of $n_c$, however, a dramatic change in the behavior of $S_R$ 
is observed.  The noise amplitude starts to increase strongly with decreasing 
$n_s$, and $\alpha$ rises rapidly from $\approx1$ to $\approx1.8$
\cite{alphaT}.  This shift of the spectral weight towards lower 
$f$ indicates a sudden and 
dramatic slowing down of the electron dynamics at $n_g\approx 10\times
10^{10}$cm$^{-2}$, which is attributed to the freezing of the electron glass.  
The same qualitative behavior, together with other manifestations of 
glassiness (slow relaxations, history dependence), was 
observed~\cite{Bogd2002} in Si MOSFETs with a much higher amount of disorder:
$\mu$ was a factor of 40 lower than in the samples studied here, and not 
surprisingly, $n_g$ was almost an order of magnitude higher.  We note that the
two sets of devices also differ substantially by their geometry, size, and 
many fabrication details (see Refs.~\onlinecite{Bogd2002,Heemskerk98}), all of
which leads us to conclude that the observed glass transition is a universal 
phenomenon in Si inversion layers, at least in those with conventional 
($d\langle\rho\rangle/dT>0$) metallic behavior~\cite{novel}.

In addition to affecting the values of $n_c$, $n_g$, and $n_{s}^{\ast}$, the 
disorder clearly plays another, nontrivial role.  In
particular, $n_c$ and $n_g$ were found to differ from each other considerably 
in low-mobility devices ($n_c$, $n_g$, and $n_{s}^{\ast}$ were 5.0, 7.5, and 
12.9, respectively, in units of $10^{11}$cm$^{-2}$) \cite{Bogd2002}, whereas 
in high-mobility devices $n_g$ is at most a few percent higher than $n_c$ [see
Figs.~\ref{expalpha}(a), (b)].  Therefore, the emergence of glassy dynamics 
here seems almost to coincide with the MIT.  Obviously, the size of the 
intermediate ($n_c<n_s<n_g$) glass phase depends strongly on disorder, in 
agreement with theoretical predictions~\cite{PastorD02}.

Earlier studies of noise in $R$ in Si MOSFETs were performed mostly at 
$T>4.2$~K and $n_s>10^{12}$cm$^{-2}$~\cite{Ralls,Kirton}.  The observed 
random telegraph and $1/f$ noise were attributed to charging and discharging
of traps in the oxide close to the Si/SiO$_2$ interface, leading to 
$dS_R/dT>0$~\cite{Rogers84}.  On the other hand, studies carried out at 
lower $n_s$ and $T$ demonstrated~\cite{Voss,Koch} clearly that the 
observed $1/f$ noise was an intrinsic property of the conduction in a 2D 
channel and not due to charge trapping.  Moreover, $dS_R/dT<0$
was found~\cite{Koch} for $T=1.5$, 4.2~K.  Our 
measurements at much lower $T$ reveal a dramatic {\em increase} of $S_R$ with 
decreasing $T$.  This rules out models of thermally activated charge 
trapping~\cite{Weissman88,DH81,Rogers84}, noise generated by 
fluctuations of $T$~\cite{VC}, and a model of noise near the 
Anderson transition~\cite{Ovadyahu92}, as possible explanations.
Likewise, the models of noise in the Mott and Efros-Shklovskii variable-range
hopping regimes~\cite{Shklovskii80} do not describe the data because they
predict either $dS_R/dT>0$ or a saturation of $S_R$ below 10-100~Hz, both in 
clear disagreement with the experiment.  Therefore, the observed noise cannot 
be a result of single electron hops even when Coulomb interactions are 
included through the Coulomb gap.  We note that no such low-frequency 
saturation of $S_R$ was found in computer simulations of a Coulomb glass, 
where $1/f$ noise was a result of transitions between many metastable states, 
with each transition being a reconfiguration of a large number of 
electrons~\cite{Kogan98}.

We have established that the exponent $\alpha\approx 1$ in the 2D metallic 
phase (above $n_g$) in both low- and high-mobility samples.  On the other
hand, $\alpha\approx 1.8$ in the glassy phase, similar to $\alpha$ found in 
some spin glasses~\cite{jjprl98,Strunk2000}, and submicron wires in the 
quantum Hall 
regime~\cite{wrobel}.  In general, such noise with spectra closer to $1/f^2$
than to $1/f$ is typical of a system far from equilibrium, in which a step 
does not lead to a probable return step.  Such high values of $\alpha$ may be
also obtained if noise results from a superposition of a small number of 
independent two-state systems (TSS)~\cite{DH81,Weissman88}.  However, even 
though some distinguishable discrete events can be seen at low $n_s$ 
[Fig.~\ref{spectra1}(a)], they do not show the characteristic repetitive form 
of stable TSS.  On the contrary, both the shape and the magnitude of noise 
exhibit random, nonmonotonic (which exclude aging) changes with time.  A 
quantitative measure of such spectral wandering is 
the so-called second spectrum $S_2(f_2,f)$, which is the power spectrum of the
fluctuations of $S_{R}(f)$ with time~\cite{Weissman93}, {\textit i.~e.} the 
Fourier transform of the autocorrelation function of the time series of 
$S_{R}(f)$. If the fluctuators ({\textit e.~g.} TSS) are not correlated, 
$S_2(f_2,f)$ is white (independent of $f_2$)~\cite{Weissman88,Weissman93} and 
equal to the square of the first spectrum.  Such noise is called Gaussian.  On
the other hand, $S_2$ has a nonwhite character, $S_2\propto 
1/f_{2}^{1-\beta}$, for interacting fluctuators~\cite{Weissman88,Weissman93}.
Therefore, the deviations from Gaussianity provide a direct probe of 
correlations between fluctuators.

\begin{figure}
\centerline{\epsfig{file=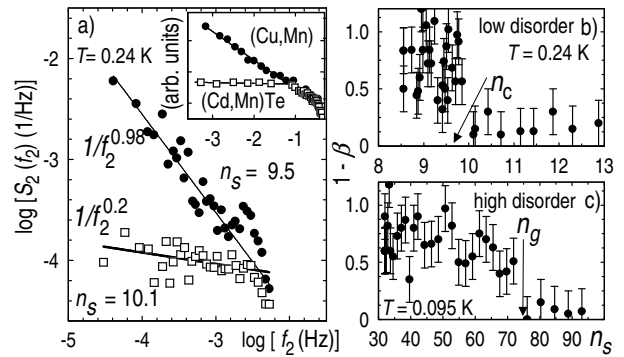,width=8cm,clip=}} \caption{(a)
Second spectral density $S_2(f_2)$ {\textit vs.} $f_2$ for $n_s 
(10^{10}$cm$^{-2})$ shown on the plot; $f_L=1$~mHz.  Inset: $\log[S_2(f_2)]$ 
for spin glasses Cu$_{0.91}$Mn$_{0.09}$ \cite{Weissman93} and 
Cd$_{0.93}$Mn$_{0.07}$Te \cite{jjprl98}.  Solid lines are fits.  Exponent 
$1-\beta$ {\textit vs.} $n_s(10^{10}$cm$^{-2})$ for (b) high-mobility and (c) 
low-mobility samples.  The low-mobility device is the same as the one studied 
in Ref.~\onlinecite{Bogd2002}.
\label{second}}
\end{figure}

We investigate $S_2$ using digital filtering~\cite{Seid96a} in a given 
frequency range $f=(f_L,f_H)$ (usually $f_H=2f_L$). 
The normalized second spectra, with the Gaussian background subtracted, are
shown in Fig.~\ref{second}(a) for two $n_s$, just above and just below $n_g$.
%
%
It is clear that there is a striking difference in the character of the two 
spectra.  Similar differences are observed between various spin glasses 
(Fig.~\ref{second}(a) inset), where $S_2$ is white~\cite{jjprl98} in the 
absence of long range interactions, and nonwhite~\cite{Weissman93} when long 
range RKKY interaction leads to hierarchical glassy dynamics~\cite{Ogielski}.
A detailed dependence of the exponent $(1-\beta)$ on $n_s$ has been determined 
for both high- and low-mobility samples (Figs.~\ref{second}(b) and (c), 
respectively).  
In both cases, $S_2$ is white for $n_s>n_g$, indicating that the observed 
$1/f$ noise results from uncorrelated fluctuators.  It is quite remarkable 
that $S_2$ changes its character in a dramatic way exactly at $n_g$ in both 
types of samples.  For $n_s<n_g$, $S_2$ is strongly non-Gaussian, which
demonstrates that the fluctuators are strongly correlated.  This, of course, 
rules out independent TSS (such as traps) as possible sources of noise when
$n_s<n_g$.  In fact, a sudden change in the nature of the fluctuators 
({\textit i.~e.} correlated {\textit vs.} uncorrelated) as a function of $n_s$
rules out 
{\em any} traps, defects, or a highly unlikely scenario that the observed 
glassiness may be due to some other time dependent changes of the disorder 
potential itself.  Instead, it provides an unambiguous evidence for the onset 
of glassy dynamics in a 2D electron system at $n_g$.

\begin{figure}
\centerline{\epsfig{file=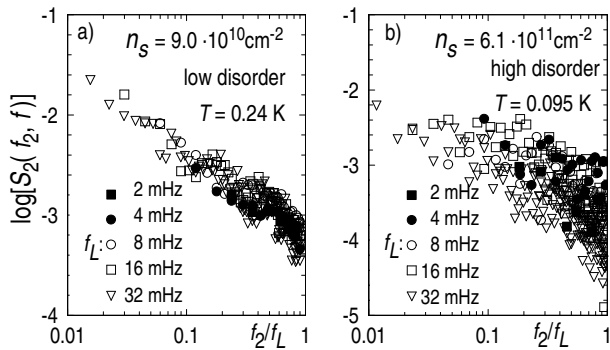,width=8cm,clip=}}
\caption{Scaling of $S_2$ measured at frequency octaves $f=(f_L,2f_L)$ for (a)
high-mobility and (b) low-mobility samples.
\label{scaling}}
\end{figure}

In the studies of spin glasses, the scaling of $S_2$ with respect to $f$ and 
$f_2$ has been used~\cite{Weissman93} to unravel the glassy dynamics and, in
particular, to distinguish generalized models of interacting droplets or 
clusters ({\textit i.~e.} TSS) from hierarchical pictures.
In the former case, the low-$f$ noise comes from a smaller number of 
large elements because they are slower, while the higher-$f$ noise comes
from a larger number of smaller elements, which are faster.  In this picture,
which assumes compact droplets and short-range interactions between them, big 
elements are more likely to interact than small ones and, hence, non-Gaussian 
effects and $S_2$ will be stronger for lower $f$.  $S_2(f_2,f)$, however, need
to be compared for fixed $f_2/f$, {\textit i.~e.} on time scales determined by
the time
scales of the fluctuations being measured, since spectra taken over a fixed
time interval average the high-frequency data more than the low-frequency data.
Therefore, in the interacting ``droplet'' model, $S_2(f_2,f)$ should be a 
decreasing function of $f$ at constant $f_2/f$.  In the hierarchical picture, 
on the other hand, $S_2(f_2,f)$ should be scale invariant: it should depend 
only on $f_2/f$, not on the scale $f$~\cite{Weissman93}.  
Fig.~\ref{scaling} shows that no systematic dependence of $S_2$ on $f$ is
seen in our samples, which
%
%
signals that the system wanders collectively between many metastable states
related by a kinetic hierarchy.  Metastable states correspond to the local 
minima or ``valleys'' in the free energy landscape, separated by barriers with
a wide, hierarchical distribution of heights and, thus, relaxation times.  
Intervalley transitions, which are reconfigurations of a large number of 
electrons, thus lead to the observed strong, correlated, $1/f$-type noise, 
remarkably similar to what was observed in spin glasses with a long-range
correlation of spin configuration~\cite{Weissman93}.  We note that, unlike 
droplet models~\cite{Fisher}, hierarchical pictures of glassy 
dynamics~\cite{Binder} do allow for the existence of a finite $T$ (or finite
Fermi energy) glass transition in presence of a symmetry-breaking field, such 
as the random potential in an electron glass.

In summary, by studying the statistics of low-$f$ resistance noise, we 
have established that the glassy ordering of a 2DES near the MIT
occurs in all Si inversion layers.  The size of the metallic
glass phase, which separates the 2D metal and the glassy insulator,
depends strongly on disorder, becoming extremely small in high-mobility 
samples.  The properties of the entire glass phase are consistent with the
hierarchical picture of glassy dynamics, similar to spin glasses with 
long-range correlations.

\begin{acknowledgments}
We are grateful to the Silicon Facility at IBM, Yorktown Heights for
fabricating low-mobility samples, and to S. Bogdanovich and V. 
Dobrosavljevi\'c for useful discussions.  This work was supported by NSF grant
DMR-0071668 and NHMFL through NSF Cooperative Agreement DMR-0084173.
\end{acknowledgments}

\end{document}